\documentclass[epj,referee]{svjour}
\usepackage{latexsym}
\usepackage{graphics}

\def\a{\alpha}
\def\b{\beta}

\begin{document}

\title{Replica-symmetry breaking in dynamical glasses}
\author{Susanna C. Manrubia\inst{1}, Ugo Bastolla\inst{1} \and 
Alexander S. Mikhailov\inst{2}}

\titlerunning{Glassy behaviour of GCLM}

\authorrunning{S.C. Manrubia, U. Bastolla, and A.S. Mikhailov}

%
%

\institute{Centro de Astrobiolog\'{\i}a, Instituto Nacional de T\'ecnica
Aeroespacial. Ctra. de Ajalvir Km.4, 28850 Torrej\'on de Ardoz, Madrid,
Spain \and Fritz-Haber-Institut der Max Planck Gesellschaft, Faradayweg 4-6,
14129 Berlin, Germany.}

\date{Received: date / Revised version: date}

\abstract{Systems of globally coupled logistic maps (GCLM) can display
complex collective behaviour characterized by the formation of synchronous
clusters. In the dynamical clustering regime, such systems possess a large
number of coexisting attractors and might be viewed as dynamical glasses.
Glass properties of GCLM in the thermodynamical limit of large system sizes
$N$ are investigated. Replicas, representing orbits that start from various
initial conditions, are introduced and distributions of their overlaps are
numerically determined. We show that for fixed-field ensembles of initial
conditions all attractors of the system become identical in the
thermodynamical limit up to variations of order $1/\sqrt{N}$, and thus
replica symmetry is recovered for $N\rightarrow \infty $. In contrast to
this, when random-field ensembles of initial conditions are chosen, replica
symmetry remains broken in the thermodynamical limit.}


\maketitle

\section{Introduction}
\label{intro}

The rich collective behaviour displayed by globally coupled logistic maps
(GCLM) \cite{Kan,Kan1} has made them to become a paradigm of complex
dynamical systems. Initially, GCLM were introduced as a mean field approach
to coupled map lattices. Subsequently, they have been used as metaphors of
neural dynamics, ecology, and cell differentiation \cite{Kbook}.
Understanding the properties of GCLM can be seen as a first step towards
grasping the dynamics and emergent properties of real, high-dimensional
systems. The dynamical equations describing the system are

\begin{equation}  \label{GCLM}
x_i (t+1) = (1 - \epsilon) f(x_i(t)) + \frac{\epsilon}{N} \sum_{j=1}^N
f(x_j(t)) \;
\end{equation}
where

\begin{equation}
f(x)=1-ax^{2}  \label{logmap}
\end{equation}
is the logistic map. The parameter $\epsilon \in [0,1]$ gives the strength
of coupling among elements. For $\epsilon =0$ the elements evolve
independently, and for $\epsilon =1$ they are synchronized already after the
first iteration and follow identical trajectories ever after. Between these
two extreme behaviours, a broad spectrum of collective dynamics emerges. The
dynamics is strongly sensitive to the control parameter $a$ of the logistic
map and depends on the size $N$ of the system. Figure 1 shows a rough phase
diagram of GCLM, based on the collective behaviour of the system which is
reached after (sometimes, very long) transients \cite{Kan,MM1,Abramson}. The
diagram includes both the parameter interval $a<1.4$ where dynamics of an
individual map is periodic and the interval $1.4<a<2$ with chaotic
individual dynamics. It contains two large domains of synchronous and
turbulent phases. They are separated by a region with glass-like behaviour.
In the synchronous domain the states of all elements are identical and the
ensemble has the same dynamics as a single map. In the turbulent phase, the
ensemble of maps is essentially desynchronized, though nontrivial
correlations have been detected even there \cite{Shibata}. The glass region
is characterized by the formation of various dynamical clusters.

Numerical simulations of GCLM have shown that, in the dynamical glass phase,
the system displays sensitivity to initial conditions. For fixed parameters $%
\epsilon $ and $a$ (and given $N$), a multiplicity of attractors can be
reached \cite{Kan,Chaos}. This property is similar to what is observed in
glassy systems, where the presence of quenched disorder causes frustration
and a large number of macroscopic configurations are possible \cite{Mezard}.
For this reason, GCLM have been described as a dynamical counterpart to spin
glasses \cite{Kanglass}, \cite{Vulpiani}. In a previous publication \cite
{MM2} two of us have introduced a replica description for this system,
defined overlaps and numerically tested replica-symmetry breaking and
ultrametric properties of GCLM. Our analysis has revealed a strong size
dependence of collective dynamics, indicating that replica symmetry might be
recovered in the thermodynamic limit $N\rightarrow \infty $. The aim of the
present paper is to investigate systematically the asymptotic statistical
properties of GCLM in this limit.

Our main result is that the asymptotic behaviour observed in the
thermodynamic limit is strongly dependent on how the ensemble of initial
conditions is prepared. In previous studies \cite{Vulpiani,MM2}, the
procedures used for random generation of initial conditions had a special
property: in the limit $N\rightarrow \infty $ all generated initial
conditions were effectively identical up to order $1/\sqrt{N}$. Therefore,
all explored attractors in the glass phase became equivalent up to
variations of order $1/\sqrt{N}$ and the replica symmetry was recovered in
the thermodynamic limit. Now we show that, if the initial conditions are
prepared in such a way that the initial field always retains macroscopic
fluctuations, the replica symmetry is clearly broken in the thermodynamic
limit $N\rightarrow \infty $. Thus, GCLM represent the first known example
of a dynamical glass with replica-symmetry breaking and, as we show, this
important statistical property does not represent a finite-size effect. Our
results also imply that the boundaries between different regions illustrated
in Fig. 1 depend on the ensemble of initial conditions, and that, for broad
ensembles and $\epsilon $ not too large, fully synchronized attractors can
coexist with multi-cluster attractors.

In the next section we introduce dynamical and statistical measures needed
to characterize clustered states of GCLM. They include the splitting
exponent, earlier proposed by Kaneko \cite{Ksplit}, and a new repulsion
exponent which we suggest. Replicas and their overlaps are defined and
attraction basin weights are considered here. In Section 3 we perform a
detailed analysis of the role of initial conditions. Replica symmetry
breaking and ultrametric properties of GCLM in the thermodynamic limit are
investigated in Section 4 under a truly random choice of initial conditions.
The paper ends with a discussion of the main results, which are compared to
the properties of other disordered systems.

\begin{figure}[tbp]
\resizebox{0.42\textwidth}{!}{\includegraphics{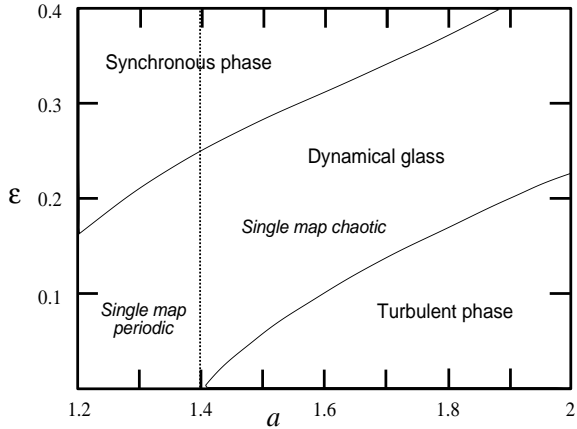}}
\caption{Rough phase space of GCLM showing the main three phases of the
system.}
\label{Fig1}
\end{figure}

\section{Characterization of attractors}

An attractor of the dynamical system (\ref{GCLM}) is characterized by the
formation of a number of clusters out of the initially symmetrical ensemble
of maps. This is one of the most intriguing properties of GCLM. Within each
cluster all elements are completely synchronized. A partition of $N$ maps
into $\mathcal{K}$ clusters is defined by indicating the numbers $N_{k}$ of
elements in each cluster, $k=1,\dots \mathcal{K}$. In the following, we
assume that the clusters have been ordered such that $N_{1}\ge N_{2}\ge
\dots \ge N_{\mathcal{K}-1}\ge N_{\mathcal{K}}$. Even if only two clusters
are present, this can correspond for $N\rightarrow \infty$ to a huge number
of different partitions, since the relative sizes of clusters may vary.

In the periodic region, and for $\epsilon =0$, the elements can be trivially
classified into $N/P$ groups, where $P$ is the period of the single map, and
elements within each group follow the periodic orbit of the single map with
different phases. At large enough $\epsilon $ the number of simultaneously
stable clusters decreases and their dynamical behavior differentiates. This
happens approximately in the whole area labeled ``dynamical glass'' in
Fig.~1. For $\epsilon $ large enough, all of the maps are synchronized and
the dynamics reduces to that of the single element (synchronous phase). Note
that fully synchronized attractors and multi-cluster attractors can coexist
in some region of parameter space.

In the glass phase, attractors of GCLM correspond to different partitions in
clusters. The global dynamics of each attractor can be periodic,
quasiperiodic or chaotic. The periodic collective dynamics is by far the
most common, and is typically found even in the parameter region where the
dynamics of a single map is chaotic.

\subsection{Splitting and repulsion exponents}

The route to synchronization can be easily found by studying how the
distance between elements evolves in time. This is ruled by the equation

\begin{equation}
x_i(t+1) - x_j(t+1) = -(1 - \epsilon)a \left(x_i^2(t)-x_j^2(t)\right) \, .
\label{shrink}
\end{equation}

Integrating it over a time $T$, one finds

\begin{eqnarray}
\left| x_i(t+T) - x_j(t+T)\right| = \exp(T\lambda_{ij}) \left| x_i(t) -
x_j(t)\right| \, , \\
\lambda_{ij}=\ln \left(a(1-\epsilon)\right)+ \frac{1}{T} \sum_t \ln
|x_i(t)+x_j(t)|\; ,  \label{shrink2}
\end{eqnarray}

If two elements belong to the same cluster $b$, their distance has to shrink
to zero. In this case, $x_i(t)\approx x_j(t)\approx X_b(t)$. Kaneko \cite
{Ksplit} defined the \textit{splitting exponent} to measure the rate of
convergence to the orbit $\{X_b(t)\}$,

\begin{equation}  \label{TE}
\lambda_b = \ln \left(2a(1-\epsilon)\right)+ \lim_{T \to \infty} \frac{1}{T}
\sum_t \ln |X_b(t)|\; ,
\end{equation}
and defined an orbit as transversely stable if it has $\lambda_b < 0$ (see
also \cite{Chaos}). While in the definition of the splitting exponent the
partition $N_1 \cdots N_{\cal K}$ does not enter, it is the distribution of the
elements into clusters which decides whether a set of orbits is a global
attractor or not. All of the stable periodic orbits of the single map have
negative splitting exponent for every positive $\epsilon$. The splitting
exponent can be positive for non-entrained elements (forming ``clusters'' of
a single element) in the chaotic domain of the logistic map.

Due to the unavoidable finite precision of digital computers, the simulation
of the deterministic system (\ref{GCLM}) can lead to pseudo-attractors which
are not stable against transversal perturbations. In the results to be
presented, we have computed $\lambda _{b}$ for all orbits and discarded
unstable attractors.

There is another condition which must be required for having a stable
partition and which, to our knowledge, has not been made explicit yet. On
the one hand, if two elements $i$ and $j$ belong asymptotically to two
different orbits $b$ and $c$, their distances (\ref{shrink2}) should remain
finite. On the other hand, since the phase space is finite, the distances
cannot diverge. Thus, the orbits of the two clusters have to fulfill the
condition

\begin{equation}  \label{TO}
\lambda_{bc} = \ln \left(a(1-\epsilon)\right)+ \lim_{T \to \infty} \frac{1}{T
} \sum_t \ln |X_b(t)+X_c(t)|=0\; .
\end{equation}
For periodic orbits, this condition is just a consequence of periodicity.
Nevertheless, it allows to rationalize some features of GCLM. We call 
$\lambda_{bc}$ \textit{repulsion exponent}, since its positive value would
mean that the two orbits repel each other, and define two orbits as 
\textit{orthogonal} if their repulsion exponent vanishes. A set of orbits is
stable if all of the orbits are transversely stable and all pairs of orbits
are orthogonal. This condition does not depend on the partition of the $N$
elements into the $\cal K$ orbits, but a precise partition is needed so that 
the set of orbits is invariant under the global dynamics.

Note that, for $\epsilon =0$, the set of all periodic orbits, stable and
unstable, fulfill the orthogonality condition, which is just equivalent to
periodicity. Thus, if the dynamics of the single map is periodic with period
$P$, the $P$ orbits obtained from the different phases of the stable
periodic orbit constitute a stable set. As $\epsilon $ increases, the
transversal stability condition becomes easier to fulfill even for lower
periodicity orbits, but the orthogonality condition becomes more difficult
(notice that the larger the number of clusters, the more demanding this
condition). Thus at some point only partitions with less than $P$ clusters
can be found. This is probably a reason why only very small numbers of
clusters are typically observed in simulations.
Additionally, since the average value $\sum_{t}\ln |X_{b}(t)+X_{c}(t)|/T$
cannot be larger than $\log (2)$, no stable two cluster system can exist for
$\epsilon >1-1/2a$. This is only an upper bound, since the actual value of 
$\epsilon $ where multiple cluster attractors disappear is much smaller.

\begin{figure}[tbp]
\resizebox{0.45\textwidth}{!}{\includegraphics{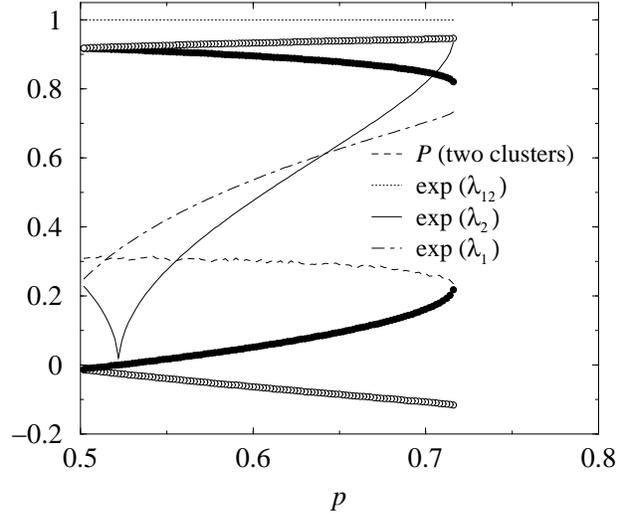}}
\caption{Possible two-cluster partitions for a system with $a=1.3$ and 
$\epsilon=0.15$. For different values of the relative size $p$ of the largest
cluster the two clusters move along two period-two orbits represented as
empty and solid circles, respectively. The fraction of initial conditions
converging to two-cluster orbits is shown as a dashed line. The remaining
trajectories synchronize completely to the stable period-four orbit. The two
transversal exponents $\lambda_1$ and $\lambda_2$ as well as the repulsion
exponent $\lambda_{12}$ are shown. Notice that $\lambda_1$ attains a minimum
at $p \simeq 0.52$. Empty circles represent the first period two orbit, full
circles represent the second one.}
\label{Fig:twoclu}
\end{figure}

In Fig.~\ref{Fig:twoclu} we show the two splitting exponents as well as the
repulsion exponent for a two-cluster system with parameters $a=1.3$ and 
$\epsilon =0.15$. There is a continuous spectrum of partitions allowed, and
for all of them the clusters move along period-two orbits, which are
periodic orbits of the two variable dynamical system
\begin{eqnarray}
&&X_{1}(t+1)=1-a\left( (1-\epsilon (1-p))X_{1}^{2}(t)+\epsilon
(1-p)X_{2}^{2}(t)\right)  \nonumber  \label{2clus} \\
&&X_{2}(t+1)=1-a\left( (1-\epsilon p)X_{2}^{2}(t)+\epsilon
pX_{1}^{2}(t)\right) \,,
\end{eqnarray}
where $p=N_{1}/N$ is the fraction of elements in the largest cluster and is
kept fixed during the dynamics. Notice that the repulsion exponent $\lambda
_{12}$ equals zero up to very high precision for all values of $p$. For
these parameter values also the completely synchronized state is stable, and
moves along a period-four attractor whose attraction basin covers roughly
70\% of phase space for the values of $p$ where the two clusters are stable.
At large $p$, the transversal exponent $\lambda _{2}$ approaches zero, and
the two-orbit system becomes unstable while the synchronized attractor
covers 100\% of phase space.

\subsection{Replicas and overlaps}

In \cite{MM2}, two of us have introduced the concept of replica in GCLM and
a measure of similarity between them, the \textit{overlap} $q^{\alpha \beta
} $. A replica $\alpha $ is an orbit $\{x_{i}^{\alpha }(t)\}$ of the whole
system, and different replicas are obtained from different realizations of
the initial conditions. Thus, the overlap $q_{\alpha \beta }$ is a random
variable dependent on the sets of initial conditions $\{x_{i}^{\alpha }(0)\}$
and $\{x_{i}^{\beta }(0)\}$. We investigate its distribution for specific
ensembles of initial conditions, keeping the parameters $\epsilon $, $a$,
and $N$ fixed.

In order to compute the overlap, we transform the orbits into binary
sequences, assigning a binary number $\sigma _{i}(t)$ to each element $i$ at
each time step $t$ such that $\sigma _{i}^{\alpha }(t)=1$ if 
$x_{i}^{\alpha}(t)>x^{*}$, and $\sigma _{i}^{\alpha }(t)=-1$ otherwise, with

\begin{equation}
x^*=\frac{-1+\sqrt{1+4a}}{2a}
\end{equation}
the fixed point of a single logistic map.

Finally, the overlap $q^{\alpha \beta}$ is defined as

\begin{equation}
q^{\alpha \beta }=\frac{1}{NT}\sum_{t=t_{0}}^{t_{0}+T}\sum_{i=1}^{N}\sigma
_{i}^{\alpha }(t)\sigma _{i}^{\beta }(t)\;.
\end{equation}
This quantity is computed after a large enough transient time $t_{0}$ has
elapsed, so that the two trajectories reach their asymptotic attractors, and
averaged over the minimal common multiple of the two periods or, in case the
asymptotic dynamics is not periodic, over a very large simulation time 
$T\simeq 10^{2-4}$, depending on the underlying dynamics. The overlap takes
values between -1 and 1. In order for it to be a meaningful measure of
similarity, the value 1 should be returned if and only if the two asymptotic
attractors coincide. Even in this case, however, the above formula can take
different values depending on the relative phase of the two orbits and on
the permutation of elements. To avoid this, the second orbit is shifted by a
number time steps with respect to the first one in order to maximize the
overlap. This procedure is similar to that used in spin glasses with
rotational symmetry \cite{Fischer}. Finally, the degeneracy due to the
arbitrary initial labelling can be avoided through a proper reordering of
the elements once the stable attractor has been reached: Elements in the
largest cluster are assigned labels from $1$ to $N_{1}$, those in the second
largest from $N_{1}+1$ to $N_{1}+N_{2}$, and so on.

Using the previous definition, the overlap $q$ returns a finite positive
value for clusters of periodic orbits and of chaotic ones in the two-band
chaos, due to the regular alternation of plus one and minus one values in
this region. Orbits with one-band chaos change the sign of the sequence 
$\sigma(t)$ in an uncorrelated fashion, implying that in the limit $T \to
\infty$ their overlap with other orbits tends to zero. Our numerical
simulations indeed show that this is the case. Thus, the definition of the
overlap becomes problematic for orbits with one-band chaos.

\subsection{Attraction basin weights}

The overlap distribution gives information on the distribution of attraction
basin weights for the particular ensemble of initial conditions chosen. In
fact, we can write it as

\begin{equation}
P(q)=\sum_{\alpha\beta} W_\a W_\b \delta(q-q_{\alpha\beta})\; ,
\end{equation}
where $\alpha$ and $\beta$ label all possible global attractors, 
$q_{\alpha\beta}$ is their overlap, and $W_\a$ and $W_\b$ are their
attraction basin weights, i.e. the fraction of initial conditions which
converge to the attractors $\alpha$ and $\beta$ respectively, and whose sum
is normalized to unity. The overlap distribution contains a delta
distribution at $q=1$, obtained for $\alpha=\beta$, whose size is equal to
the average attraction basin weight:

\begin{equation}
Y=\sum_{a}W_{a}^{2}\;.  \label{Y}
\end{equation}
This parameter expresses the probability that two randomly chosen initial
conditions converge to the same attractor and is able to distinguish between
different situations. If $Y$ is equal to or tends to one in the
thermodynamic limit, it means that there is only one relevant attractor
which attracts in this limit all of the phase space of the system. If $Y$
tends to zero in the thermodynamic limit, it means that the system has in
this limit a diverging number of different attractors and none of them is
dominant. The situation in between, when $Y$ is finite but smaller than one,
means that there is a finite number of relevant attractors.

For a fixed value of $a$ and increasing $\epsilon $, the collective
behaviour of GCLM changes from turbulent to glassy (multiple clusters) to
finally fall into complete synchronization. This last transition can be
characterized through different parameters. Kaneko proposed to use the
average cluster number \cite{Kan}, which grows continuously from a finite
value in the dynamical glass phase to unity in the synchronous phase (for
fixed $N$). An alternative measure can be the fraction $W_{\mathrm{CS}}$ of
initial conditions converging to the single coherent attractor. It turns out
that multiple-cluster attractors have vanishing attraction basins in the
thermodynamic limit, so that the only nonvanishing contribution to $Y$ comes
from the completely synchronized attractor, and we can approximate $Y\simeq
W_{\mathrm{CS}}^{2}$. Thus we can also study through the parameter $Y$ the
transition between complete and partial synchronization.

\section{Fixed-field ensemble}

\label{sec3}

To compute overlap distributions and other statistical properties of GCLM, a
set of replicas corresponding to different initial conditions should be
taken. Ideally, \textit{all} initial conditions should be present in the
set. In an actual computation, this is never possible. Instead, a large
ensemble of initial conditions is randomly generated. One expects that
averaging over this ensemble would be equivalent to the ideal averaging over
``all'' initial conditions. However, this would only be true if the employed
random set is relatively uniformly sampling the full space of initial
conditions. Some random sets of initial conditions may be missing this
property. In previous numerical investigations \cite{Vulpiani,MM2} of
dynamical glasses, to generate a set of initial conditions the coordinate 
$x_{i}(0)$ of each map $i=1,...,N$ was chosen independently and with a
uniform probability density from the interval $(-1,1)$. It was tacitly
assumed that this procedure would yield uniform sampling of initial
conditions. However, the initial conditions generated in this way become
increasingly similar in the thermodynamic limit $N\to \infty $.

As follows from (\ref{GCLM}) and (\ref{logmap}), the dynamics of GCLM is
described by the equations
\begin{equation}
x_{i}(t+1)=1-a\left[ (1-\epsilon )x_{i}^{2}(t)+\epsilon m(t)\right]
\label{firststep}
\end{equation}
where
\begin{equation}
m(t)=\frac{1}{N}\sum_{i=1}^{N}x_{i}^{2}(t)  \label{glob}
\end{equation}
is the \textit{synchronizing field} that acts on a given element $i$ and is
collectively produced by the whole system.

Let us consider statistical properties of the initial synchronizing field 
$m=m(t=0)$ in the limit of large $N$, when the coordinates $x_{i}(0)$ of each
map $i=1,...,N$ are chosen independently and with a uniform probability
density from the interval $(-1,1)$. Because this field represents then a sum
of a large number of independent random variables, it should obey for $N\to
\infty $ a Gaussian probability distribution
\begin{equation}
p(m)=\frac{1}{\sqrt{2\pi \sigma }}\exp \left[ -\frac{\left( m-\overline{m}
\right) ^{2}}{2\sigma }\right]  \label{Gauss}
\end{equation}
where $\overline{m}$ is the mean value of the field $m$ and $\sigma $ is its
mean-square statistical variation. Using (\ref{glob}), we obtain
\begin{equation}
\overline{m}=\frac{1}{N}\sum_{i=1}^{N}\langle x_{i}^{2}(0)\rangle =\frac{1}{
2N}\sum_{i=1}^{N}\int_{-1}^{1}x^{2}dx=\frac{1}{3}  \label{mean}
\end{equation}
and
\begin{equation}
\sigma =\langle (m-\overline{m})^{2}\rangle =\frac{1}{N^{2}}%
\sum_{i=1}^{N}(\langle x_{i}^{4}(0)\rangle -\langle x_{i}^{2}(0)\rangle
^{2})=\frac{4}{45N}  \label{disp}
\end{equation}

Thus, in the thermodynamic limit $N\to \infty $ the initial synchronizing
field approaches a constant value, independent of the realization. For large
$N$'s it shows fluctuations of order $1/\sqrt{N}$. Such a set of initial
conditions shall be called a \textit{fixed-field ensemble }below.

Below in this section we discuss glass properties of GCLM for evolutions
starting from a fixed field ensemble (to our knowledge, this is the only way
in which the initial conditions have been so far modelled in the
literature). As we shall see, this ensemble leads to replica symmetry, since
all approached attractors are then identical up to small variations which
vanish in the thermodynamic limit.

\subsection{Transition to complete synchronization}

At sufficiently high coupling strength $\epsilon $, GCLM become completely
synchronized. We characterize the transition to complete synchronization
through the probability $Y=P(q=1)$ (see Eq.~(\ref{Y})) that two randomly
chosen initial conditions from a fixed-field ensemble fall into the same
attractor, see Fig.~3. Our analysis is performed with a value of the
logistic parameter $a=1.3$ where the single logistic map is periodic with
period four. The coexistence of different attractors and their dynamics for
this parameter choice have been previously considered in \cite{Chaos}. Near
the transition to complete synchronization the final stable attractors
consist of two clusters following the dynamics of period two.

\begin{figure}[tpb]
\label{Fig:Ktrans} 
\resizebox{0.45\textwidth}{!}{\includegraphics{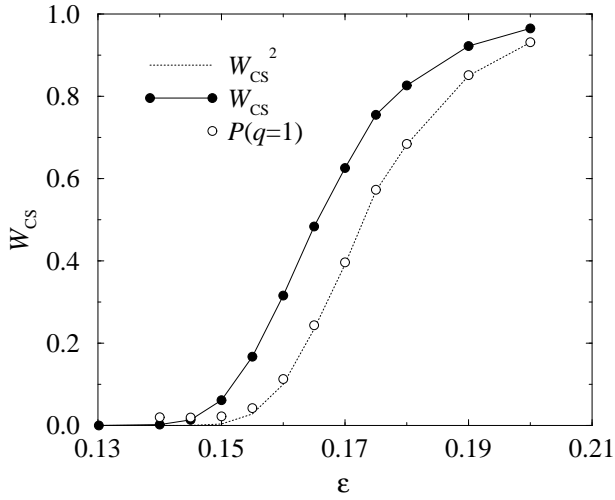}}
\caption{Transition to complete synchronization. When coupling 
$\epsilon$ is increased, the attraction basin of the completely
synchronous state grows until all initial conditions end there.}
\end{figure}

\begin{figure}[tbp]
\label{P1trans} 
\resizebox{0.45\textwidth}{!}{\includegraphics{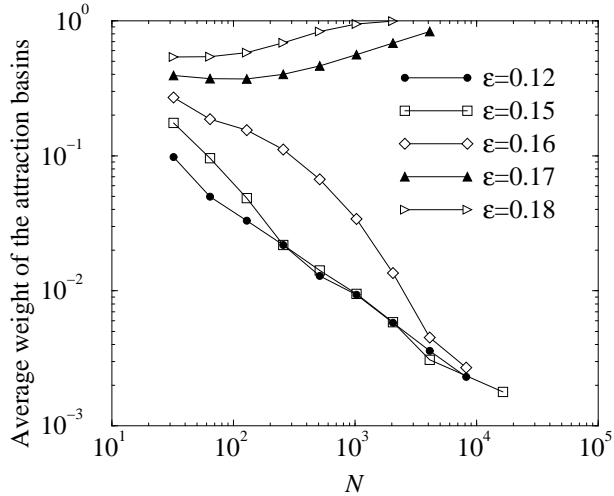}}
\caption{Average weight of an attractor basin $Y=P(1)$ as a function 
of the system size for $a=1.3$.}
\end{figure}

It is interesting how this transition takes place in the thermodynamic
limit. Each curve in Fig.~4 represents $Y$ for a given value of $\epsilon $
as a function of $N$. Even if the completely synchronized state is stable
for $\epsilon >\epsilon _{1}(a)$, the smallest value at which the
synchronized state is transversally stable, this state is never observed
until $\epsilon $ reaches a larger value $\epsilon > \epsilon_{2}(a)$ \cite
{Chaos}. The transition is discontinuous (first order) and takes place at $%
\epsilon \simeq 0.165$ for $a=1.3$. For smaller $\epsilon$ the system is in
a phase where many different attractors coexist. All of them are two-cluster
attractors. Since the average attraction basin weight $Y$ vanishes in the
thermodynamic limit, the number of different attractors diverges for $%
N\rightarrow \infty $. The unbounded increase of the number of different
attractors for $N\to \infty$ was indeed one of the first indications that
GCLM might represent a glass-like system \cite{Vulpiani}.

We have further examined how the number of attractors $\mathcal{M}$ visited
by the system grows as the number $\mathcal{I}$ of different initial
conditions used increases. Our results for $\epsilon =0.15$ and $a=1.3$ are
displayed in Fig.~5. We observe an approximate power-law dependence $%
\mathcal{M}\propto \mathcal{I}^{\eta }$, with an exponent $\eta $ dependent
on the system size $N$. For $\mathcal{I\to \infty }$, $\mathcal{M}$ should
saturate at a finite value. Although the number of different attractors
grows fast, the bending at large $\mathcal{I}$ for the largest size reflects
the existence of an asymptotic value $\mathcal{\ M}_{\infty }(N)$. 

\begin{figure}[tbp]
\label{attic}\resizebox{0.45\textwidth}{!}{\includegraphics{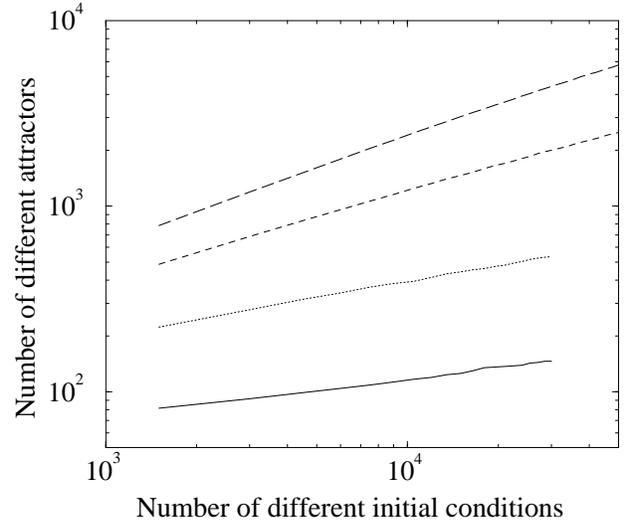}}
\caption{The number of different attractors reached by the system increases
as a power law of the number of random initial conditions used. Increasing
the latter number is equivalent to exploring in higher detail the phase space.
From top to bottom, $N=16384$, $4096$, $1024$, and $256$. }
\end{figure}

\subsection{Distributions of overlaps}

\begin{figure}[tbp]
\resizebox{0.45\textwidth}{!}{\includegraphics{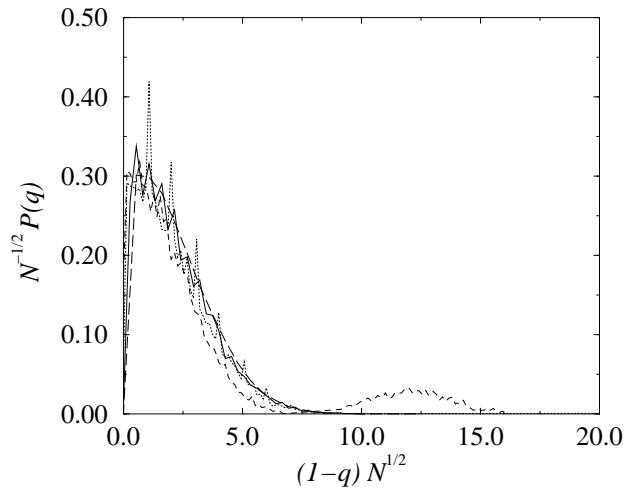}}
\caption{Data collapse of the normalized overlap distribution under increasing
system size; system parameters $\epsilon=0.15$ and $a=1.3$. }
\end{figure}

To quantify the similarity between different attractors, we have calculated
the overlap distributions for the same parameters, $\epsilon =0.15$ and $%
a=1.3$, and different system sizes $N$. As seen in Fig.~6, such distribution
$P(q)$ approaches a delta-function in the thermodynamic limit $N\rightarrow
\infty $. The width of the main peak in $P(q)$ goes to zero as $1/\sqrt{N}$.
Some finite-size effects can be observed in the bump at the smallest size
represented, and in the peaks which appear intermittently for relatively
large values of $N$, showing the ``locking'' of the system close to prefered
partitions, before reaching the asymptotic behaviour.

The finite-size behaviour can also be more complicated. Fig. 7 shows the
overlap distributions obtained at the same control parameter $a=1.3$ for
four different values of $\epsilon $ and systems of size $N=256$. For small $%
\epsilon $ (Fig.~7a), there is a large number of partititions close to the
attractor with the largest basin, $N_{1}=149$, $N_{2}=107$. A second group
of attractors corresponds to three-cluster families close to $N_{1}=115$, $%
N_{2}=85$, $N_{3}=55$. The attractors within each group are similar and
their mutual overlaps are close to unity, explaining the large weight of $%
P(q)$ at $q\simeq 1$. The overlaps between these two groups give a second
contribution around $q=0.7$. The continuous line in Fig.~7a shows the total
distribution, the dashed line corresponds to the attractors with the same
number of clusters, and the dotted line represents the contribution from the
overlaps between two- and three-cluster attractors. The part of the
distribution close to $q=1$ results from three-cluster attractors where the
third cluster has only a few elements. As the coupling strength grows, the
three-cluster attractors become less and less frequent, and two-cluster
attractors dominate (Fig.~7b,c). For large enough coupling (an example is $%
\epsilon =0.17$ in Fig.~7d), the completely synchronous state appears and
starts to occupy an increasingly large fraction of the phase space. Its
self-overlap is unity, while its overlap with the remaining two-cluster
attractors is small (three-cluster attractors are no longer present). The
overlap between one- and two-cluster attractors explains the large
contribution at small values of $q$ observed in Fig.~7d. If $\epsilon $
increases further, the completely synchronous state attracts more and more
initial conditions, $P(q)$ tends to a delta-function at $q=1$, and the
contribution at small $q$ disappears.

\begin{figure}[tbp]
\resizebox{0.45\textwidth}{!}{\includegraphics{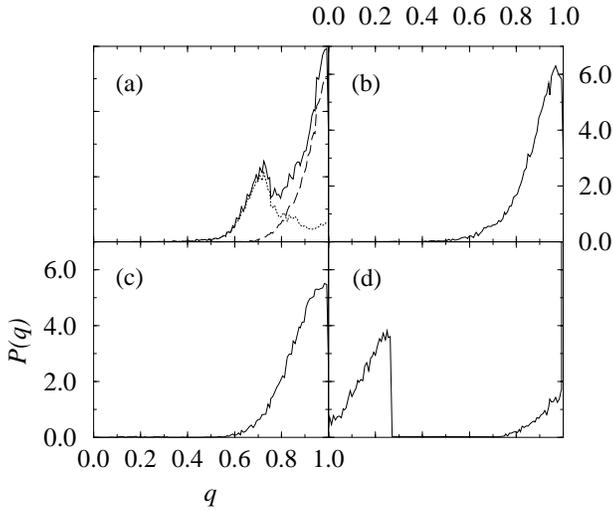}}
\caption{Overlap distributions for $N=256$ and $a=1.3$. (a) $\epsilon=0.05$,
(b) $\epsilon=0.1$, (c) $\epsilon=0.14$, and (d) $\epsilon=0.17$. }
\end{figure}

In the chaotic domain of a single logistic map, for $a>a_{\infty }$, a
similar behavior but with strong finite-size effects is observed. In a
previous publication \cite{MM2}, we have studied the parameters $\epsilon
=0.1$ and $a=1.55$. This point had been also analysed in \cite{Vulpiani} for
its glassy properties. The overlap distribution is broad here and its shape
keeps almost unchanged with the system size until $N\simeq 4000$. But when $%
N $ increases further, the most of the distribution's weight is shifted
towards $q\simeq 1$, indicating that the attractors reached by the system
indeed become very similar. For $\epsilon =0.3$ and $a=1.9$, we have
observed that for sizes up to $N\simeq 3000$, the overlap distribution $P(q)$
remains almost constant (see Fig.~4 in \cite{MM2}). If we apply rescaling
similar to Fig.~6, only the two largest sizes ($N=2048$ and $N=8192$) seem
to follow the expected asymptotic behaviour and collapse.

Our analysis shows that, for $N$ large enough, GCLM tend to a prefered
cluster partition. It can be said that the same macroscopic state is always
found in the thermodynamic limit, apart from ``thermal fluctuations'' of
order $1/\sqrt{N}$. The origin of such fluctuations lies in the $1/\sqrt{N}$
variations of the synchronizing field.

\subsection{Distributions of cluster sizes}

Information similar to the overlap distribution is contained in the
distribution $Q(p_{k})$ of cluster sizes $p_{k}=N_{k}/N$, i.e. of the
fractions of elements belonging to a cluster (obviously, $\sum_{k}p_{k}=1$).
For large sets of initial conditions (varying between $10^{4}$ and $2\times
10^{5}$ realizations) in the fixed-field ensemble, we have computed the
values of $p_{k}$ for all stable partitions and thus obtained the
distributions $Q(p_{k})$ of cluster sizes.

An example of such distributions for $\epsilon =0.15$ and $a=1.3$ and
different system sizes is shown in Fig.~8a. We see that in this case the
asymptotic attractors are always formed by two clusters of unequal size.
Their dynamics is periodic with period two. As $N\to \infty $, the
distribution $Q(p_{k})$ shrinks in width around the two prefered sizes, $%
p_{1}\simeq 0.372$ and $p_{2}=1-p_{1}$. For large enough $N$ the two peaks
approach a Gaussian distribution, and its width decreases proportional to $1/%
\sqrt{N}$ , as shown in Fig.~8b.

\begin{figure}[tbp]
\resizebox{0.45\textwidth}{!}{\includegraphics{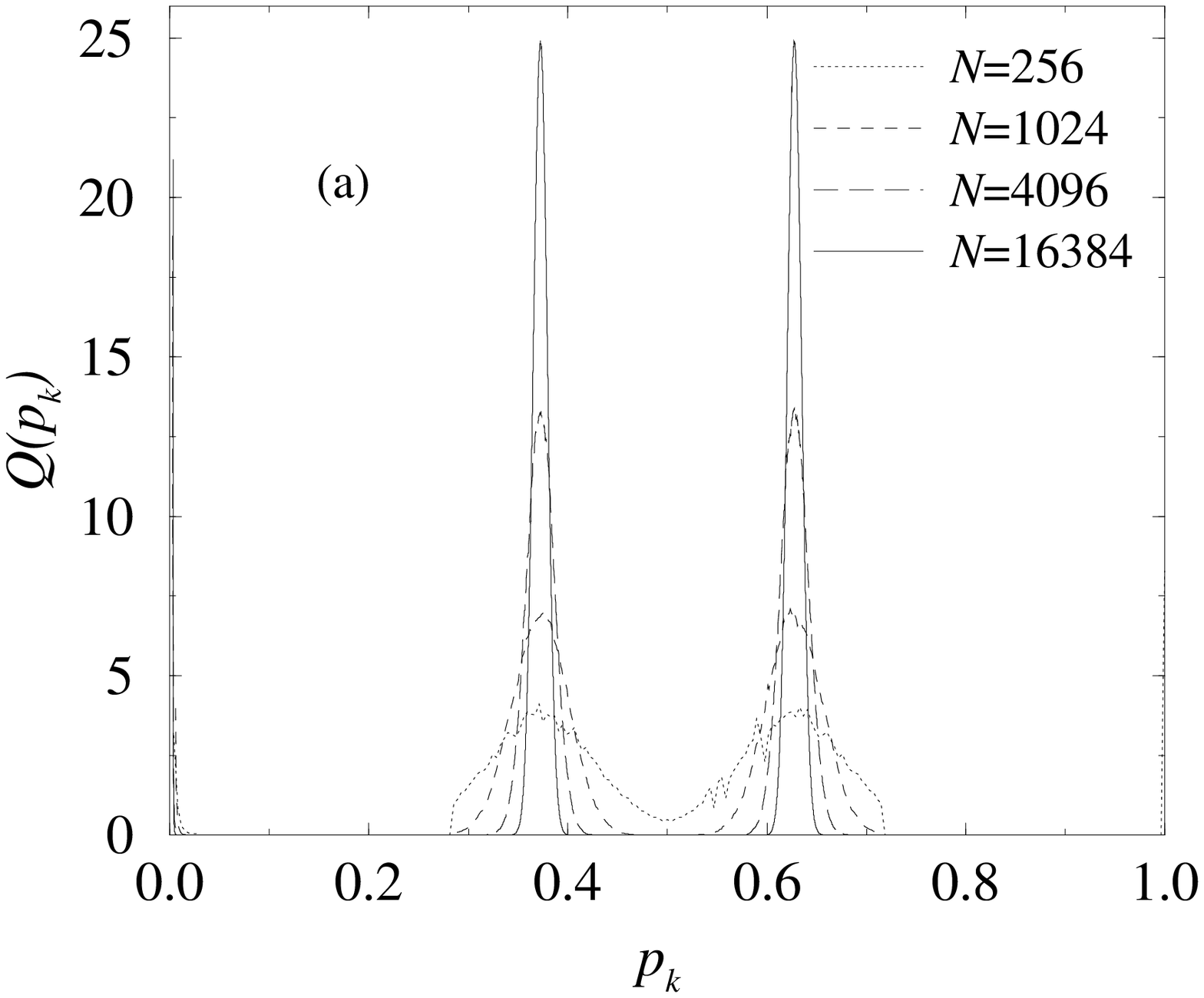}} 
\resizebox{0.45\textwidth}{!}{\includegraphics{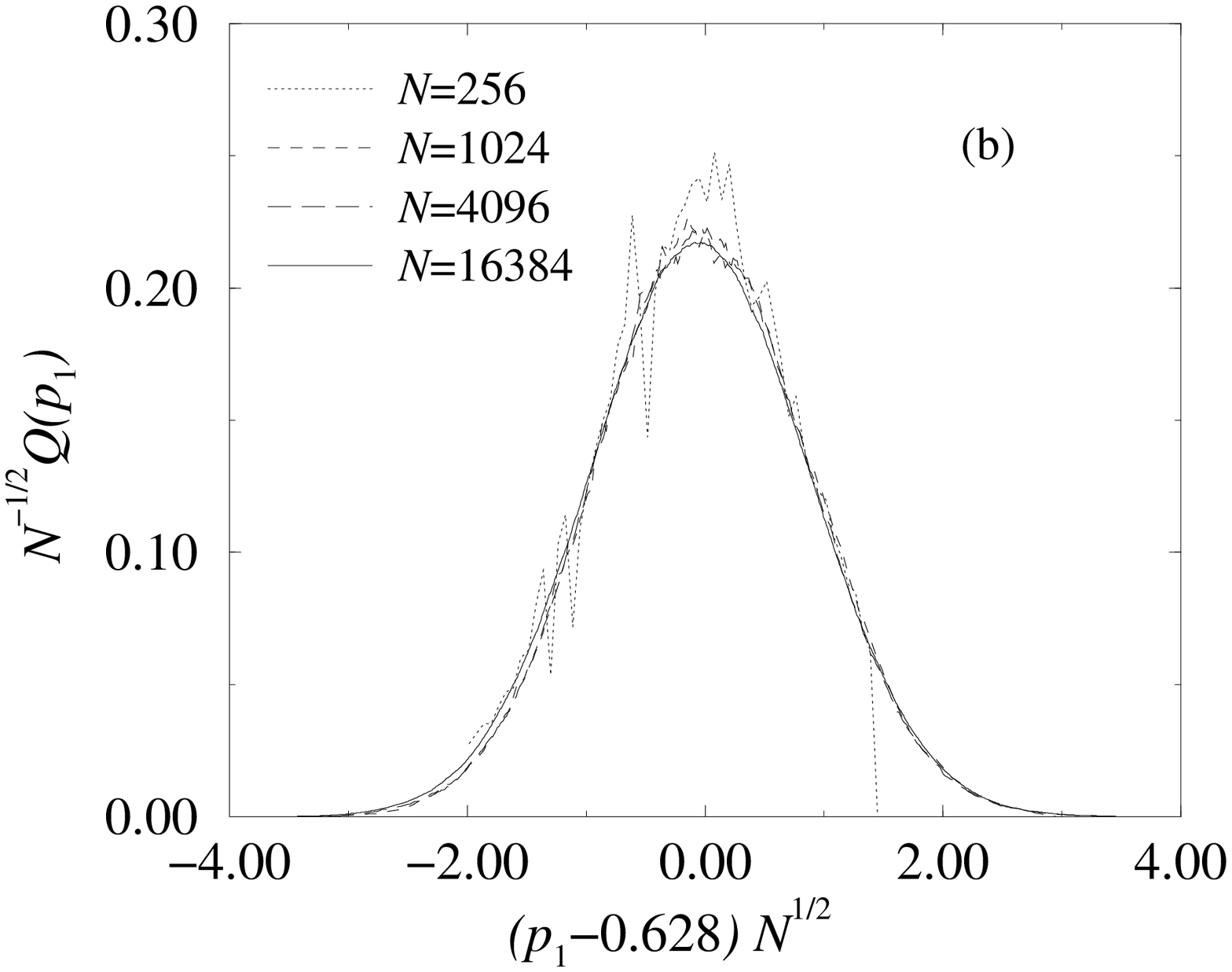}}
\caption{(a) Distribution of cluster sizes for increasing system size;
parameters $\epsilon=0.15$ and $a=1.3$. (b) Data collapse of the size
distribution for the largest cluster, with $N_1 \simeq 0.628 N$.}
\end{figure}

A similar behaviour was observed for other parameter values. Generally, for
sufficiently large $N$ the distribution of the sizes of the $k$th cluster is
a Gaussian centered at a prefered value $p_{k}^{*}$. We show two more
examples. In Fig.~9a ($\epsilon =0.3$ and $a=1.9)$, the system tends to
attractors with two clusters of almost equal sizes, though occasionally also
three-cluster attractors are observed. Fig.~9b ($\epsilon =0.1$ and $a=1.55$%
) gives an example where attractors with four clusters of different sizes
are prefered. We show the total size distribution $Q(p_{k})$ together with
the size distributions for clusters of rank one to four. Here, the prefered
partition is close to $p_{1}=0.307$, $p_{2}=0.242$, $p_{3}=0.2295$, and $%
p_{4}=0.2215$.

Thus, we have investigated numerically the statistical properties of GCLM in
the glass phase starting from initial conditions in the fixed-field
ensemble. The investigations show that in the thermodynamic limit $%
N\rightarrow \infty $ this system has a great number of different
attractors, increasing as a power law of system size $N$. However, all these
attractors are very similar. Namely, the differences in their statistical
properties, such as the cluster sizes, are proportional to $1/\sqrt{N}$ and
thus vanish in the limit $N\rightarrow \infty $. This explains why
replica-symmetry is recovered in the thermodynamic limit, when only the
evolutions initiating from a fixed-field ensemble are considered.

\begin{figure}[tbp]
\resizebox{0.45\textwidth}{!}{\includegraphics{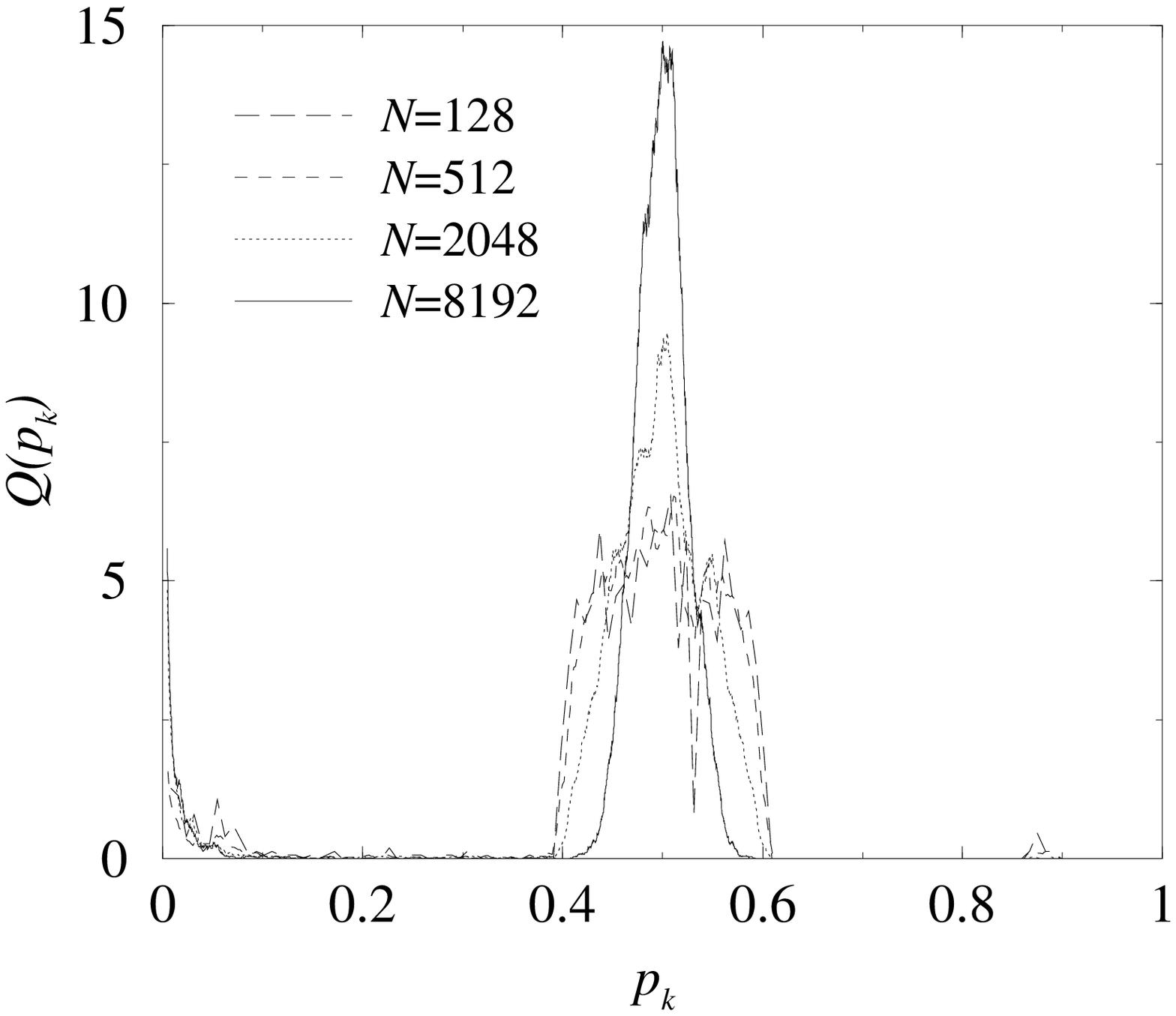}} %
\resizebox{0.45\textwidth}{!}{\includegraphics{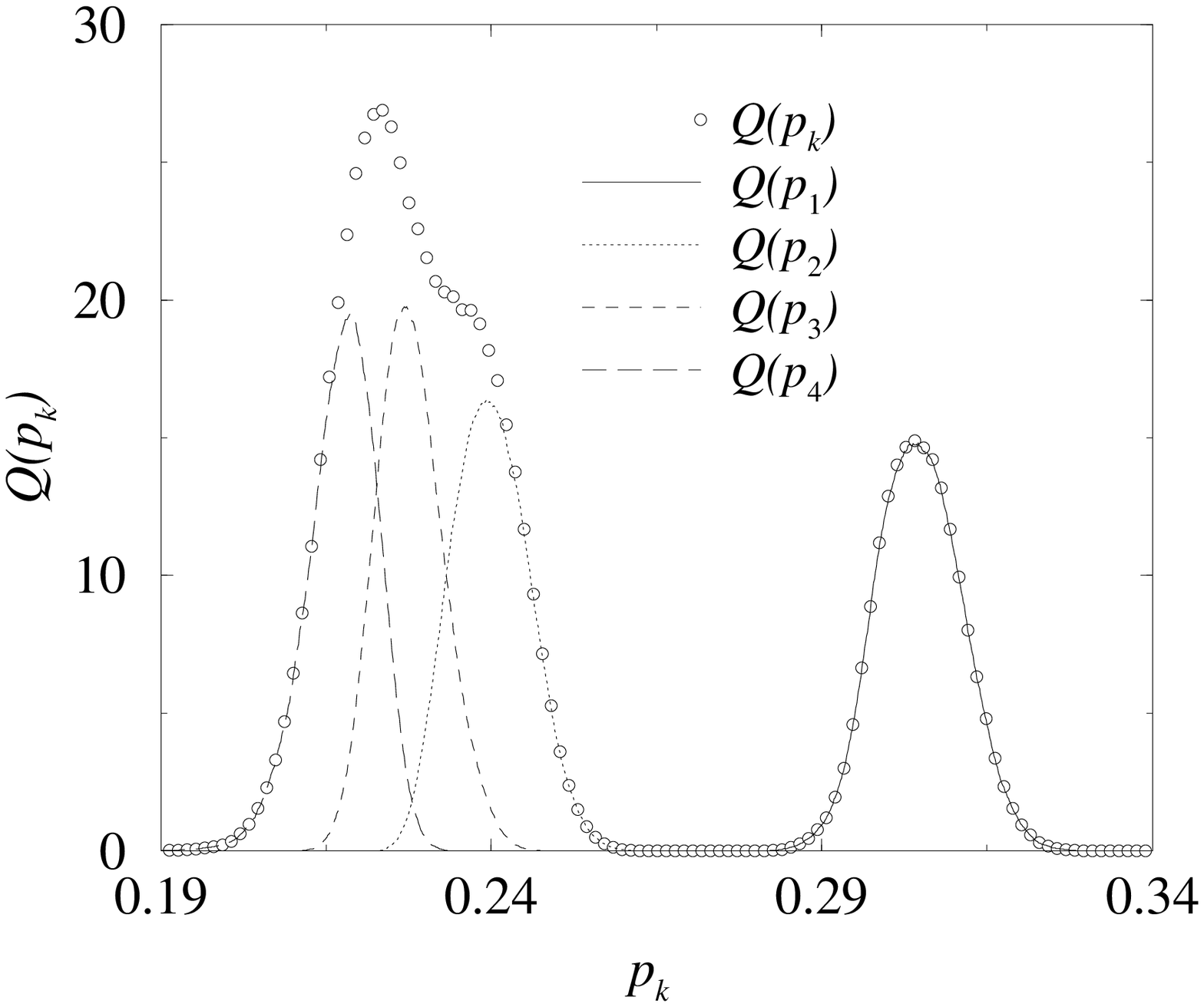}}
\caption{(a) Distribution of cluster sizes for different system sizes, $%
\epsilon=0.3$, and $a=1.9$. For $N$ large enough, the system chooses a
partition formed by two clusters of similar sizes. (b) Distribution $Q(p_k)$
for $N=8192$ and parameters $\epsilon=0.1$ and $a=1.55$. The
whole distribution $Q(p_k)$ and individual distributions for the
largest, second, third, and fourth largest clusters are displayed.}
\end{figure}

\section{Random-field ensemble}

\label{sec4}

\subsection{The role of initial conditions}

For given parameter values $a$ and $\epsilon $, many different orbits
corresponding to a continuous spectrum of two-cluster partitions and to
complete synchronization are stable (see Fig.~\ref{Fig:twoclu}). Yet, only
one partition is chosen in the fixed-field ensemble, up to variations
of order $1/\sqrt{N}$. The selection of this prefered partition cannot be
explained by a higher stability of its orbit. For instance, Fig.~\ref
{Fig:twoclu} indicates that for $a=1.3$ and $\epsilon =0.15$ the transversal
stability is strongly increased at the cluster size $p_{1}\simeq 0.522$.
However, the selected partition in the fixed-field ensemble has in this
case the cluster size $p_{1}=0.628$ whose stability is much weaker. For
parameters in the chaotic domain of a single logistic map, the situation is
similar. Stable attractors with chaotic dynamics exist here, but the system
often shows a preference for the periodic ones. 

To examine in more detail the role of initial conditions, we use a slightly
generalized version of the fixed-field ensemble. Namely, we assume now that
the initial coordinates $x_{i}(0)$ of all maps are independently and
uniformly distributed between $-\xi $ and $\xi $ (note that the fixed-field
ensemble corresponds then to the choice $\xi =1$). Calculating again the
statistical distribution of initial synchronizing fields $m$, we find that
in the thermodynamic limit $N\rightarrow \infty $ it is again given by a
Gaussian distribution with mean value $\overline{m}=\xi /3$ and mean-square
dispersion $\sigma =(4/45)(\xi /N)$. Hence, at time $t=1$ the coordinates $%
x_{i}(1)$ of the maps are given by

\begin{eqnarray}
x_{i}(1) &=&1-a\left[ (1-\epsilon )x_{i}^{2}(0)+\epsilon m\right]  \nonumber
\\
&=&1-a\left[ (1-\epsilon )x_{i}^{2}(0)+\epsilon {\frac{\xi }{3}}%
+O(N^{-1/2})\right] \;.
\end{eqnarray}

\begin{figure}[tbp]
\resizebox{0.42\textwidth}{!}{\includegraphics{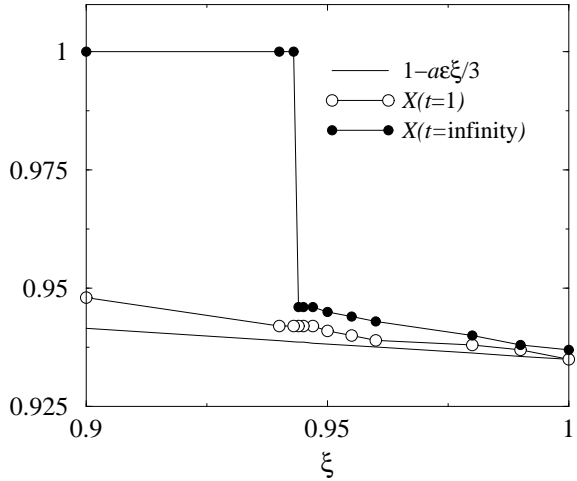}}
\caption{For different values of the initial field, the figure shows the
most likely value of $x_i(1)$ (empty circles) and the coordinate of the
prefered attractor closest to this value (full circles). For $\xi<0.944$
the prefered attractor is completely synchronized.}
\label{Fig:ini}
\end{figure}

Since $x_{i}(0)$ is uniformly distributed, $x_{i}^{2}(0)$ is distributed
with density $1/2x$ which diverges at $x_{i}^{2}(0)=0$, so that the
distribution of $x_{i}(1)$ has a pronounced maximum at $x=1-a\epsilon \xi /3$
. This initial bias drives the system towards the attractor closest to the
most probable value of $x_{i}(1)$. This is shown in Fig.~\ref{Fig:ini},
where we represent, as a function of the fixed ensemble parameter $\xi $,
the value $1-a\epsilon \xi /3$ where the maximum of $x_{i}(1)$ is expected,
the actually observed maxima of $x_{i}(1)$, and the coordinate on the
prefered asymptotic orbit. Varying $\xi $, the bias in the initial value $%
x_{i}(1)$ changes and drives the system to different asymptotic orbits (full
circles in Fig.~\ref{Fig:ini}), in turn corresponding to different
partitions of the elements. We thus find $p=0.632$ at $\xi =1$ and $p=0.717$
at $\xi =0.944$. Partitions with larger $p$ cannot lead to two transversely
stable orbits, thus for $\xi <0.944$ the completely synchronized attractor
is always reached.

\begin{figure}[tbp]
\resizebox{0.42\textwidth}{!}{\includegraphics{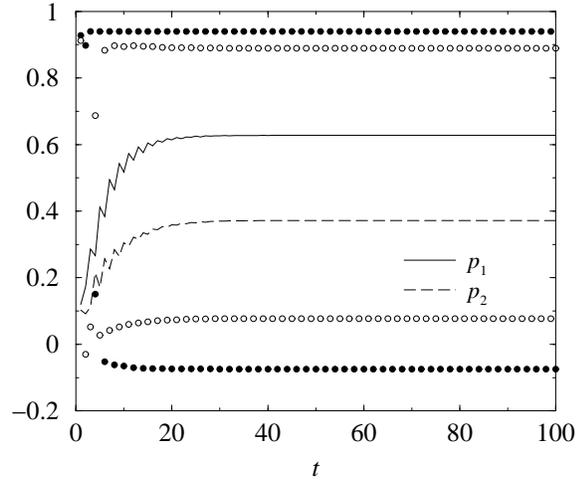}}
\caption{Dynamics of two-cluster synchronization for $a=1.3$, $\epsilon=0.15$
and $N=1000$. The solid circles represent the coordinates of the largest
cluster, empty circles those of the second larger cluster, the solid and
dashed lines represent their sizes respectively.}
\label{Fig:map1}
\end{figure}

It is interesting, in this framework, to look at the dynamics of
synchronization for $a=1.3$, $\epsilon =0.15$ and varying $\xi $ (see Fig.~%
\ref{Fig:map1}). Different examples always show the same pattern: after the
first time step, the most populated region of phase space coincides with the
maximum in the distribution of $x_{i}(1)$. After very few time steps, the
two most populated ``clusters'' start to oscillate close to a stable
attractor made of two periodic orbits of period two, but most elements are
not synchronized yet and the partition is quite different from the final
one. At the same time as oscillations around the periodic orbits are dumped,
more and more elements join the two clusters, until the partition which
stabilizes the periodic orbits is reached. Thus the system first chooses the
orbits and only afterwards partitions which would stabilize them.

\begin{figure}[tbp]
\resizebox{0.42\textwidth}{!}{\includegraphics{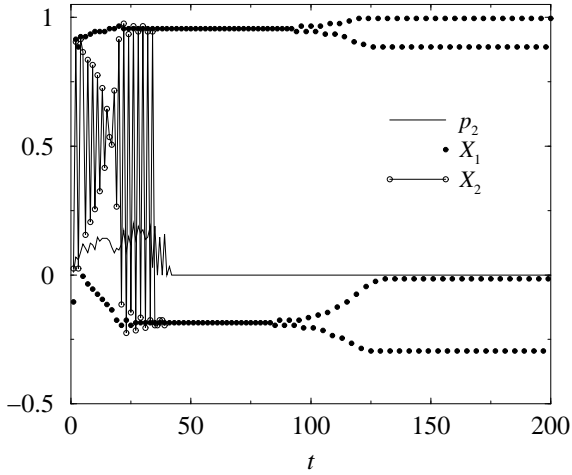}}
\caption{Dynamics of complete synchronization for $a=1.3$, $\epsilon=0.2$
and $N=1000$. Solid circles represent the coordinates of the largest
cluster, empty circles (linked by a line to guide the eye) those of the
second larger cluster, and the solid line its size $p_2$. The second cluster
is attracted by the first one and disappears at $t\simeq 40$, but the system
continues oscillating on a metastable period-two orbit until, through a
dynamical bifurcation, it reaches the stable period-four orbit.}
\label{Fig:map5}
\end{figure}

\begin{figure}[tbp]
\resizebox{0.42\textwidth}{!}{\includegraphics{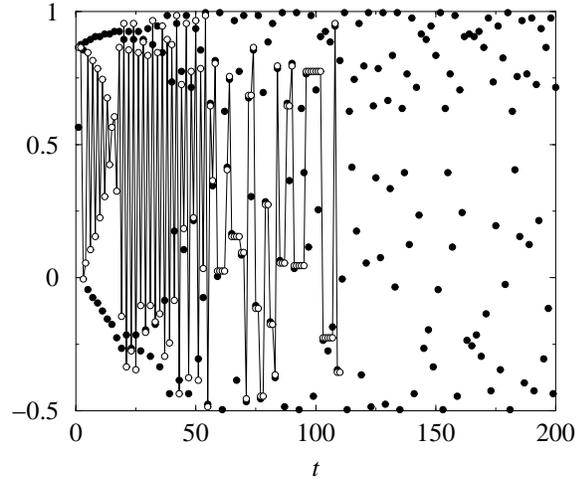}}
\caption{Dynamics of complete synchronization for $a=1.3$, $\epsilon=0.25$
and $N=1000$. Solid circles represent coordinates of the largest
cluster, empty circles those of the second larger cluster. The second
cluster is attracted by the first one and disappears at $t=110$. The
asymptotic dynamics is chaotic, but the largest cluster approaches a
metastable period-two orbit in the first stage of the dynamics.}
\label{Fig:map7}
\end{figure}

A similar route is observed even when the system tends to the completely
synchronous period-four orbit. First the system approaches two period-two
orbits which are very close one to each other and starts to partition on
them. At some point, the smaller cluster is attracted by the larger one and
disappears. For some parameter values the period-two orbit remains
metastable for quite a long time, until it splits into a period-four orbit
through a kind of dynamical bifurcation (see Fig.~\ref{Fig:map5}). Even for
values of $a$ in the chaotic phase we observed synchronization first through
attraction towards prefered period-two orbits and then through a
bifurcation (see Fig.~\ref{Fig:map7}).

Summaryzing the findings of this section, we can say that the initial
synchronization field strongly biases the elements towards a prefered
region of phase space, leading them to periodic orbits which can be either
stable (and indeed stabilized through the appropriate partition of the
system) or metastable (and eventually transformed to completely synchronous
attractors).

\subsection{Replica-symmetry breakdown in the random-field ensemble}

As shown in the previous section, by varying the parameter $\xi $ we can
drive the system to macroscopically different attractors. Thus an ensemble
of initial conditions, where $\xi $ is randomly chosen for each initial
state, is expected to lead to very different attractors. We define a \textit{%
random-field ensemble} as a random set of initial conditions which is
generated in the following way: For each realization, we first choose at
random the parameter $\xi $ form the interval (0,1). Then the initial states
$x_{i}(0)$ of all individual maps in the system are independently drawn
from the interval $(-\xi ,\xi ).$
\begin{figure}[tbp]
\resizebox{0.45\textwidth}{!}{\includegraphics{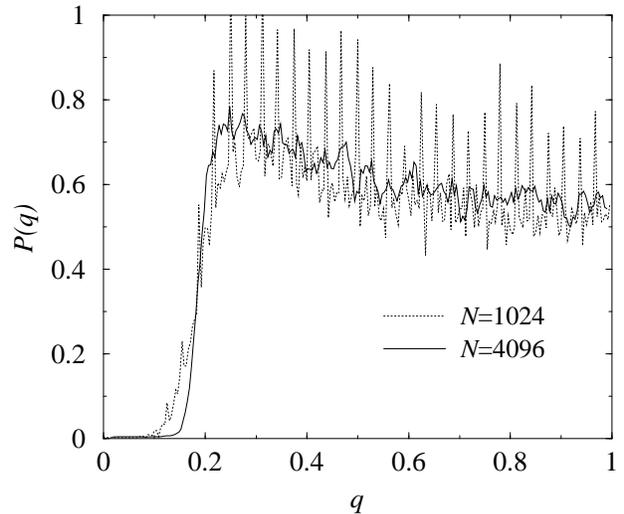}}
\caption{Distributions of overlaps for $a=1.3$ and $\epsilon =0.15$ in the
random-field ensemble for different system sizes $N$. The value of 
$P(1)$ is not represented. It accumulates around 50\% of the total 
weight of $P(q)$.}
\label{Fig:contover}
\end{figure}

\begin{figure}[tbp]
\resizebox{0.45\textwidth}{!}{\includegraphics{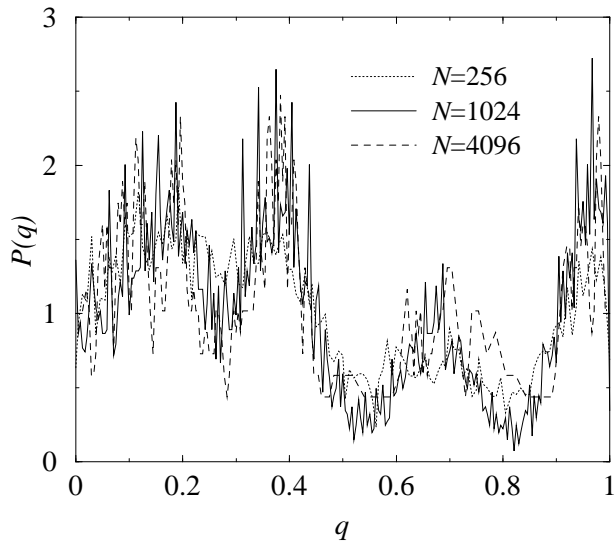}}
\caption{Distributions of overlaps for $a=1.55$ and $\epsilon=0.1$ in the
random-field ensemble for different values of $N$. For these 
parameters the single map is chaotic. The value of $P(1)
\simeq 0$ for all $N$.}
\label{Fig:contover2}
\end{figure}

When such random-field ensembles of initial conditions are used, overlap
distributions do not shrink to a delta-function peak, but remain continuous
in the thermodynamic limit. This is shown in Fig.~\ref{Fig:contover}, which
displays overlap distributions in the random-field ensemble for $a=1.3$ and $%
\epsilon =0.15$ and different system sizes. In this case, the weight of the
completely synchronized attractor, $Y=P(q=1)$ does not vanish in the glassy
phase. We expect that the transition to complete synchronization is in this
case second-order like: $Y$ tends continuosly to unity as $\epsilon $
approaches the critical coupling at which only synchronized orbits are
stable.

We present also the distributions of overlaps for parameters in the chaotic
domain of the single map, $\epsilon =0.1$ and $a=1.55$, and three different
values of the system size $N$. There is again a qualitative difference
between the fixed-field and the random-field ensemble. While in the former
case the function $P(q)$ showed a systematic loss of structure for
increasing $N$ (see Fig.~1 in \cite{MM2}), in the latter situation it
remains remarkably invariant with the growth of the system size.

Thus, we see that for the random-field ensembles of initial conditions the
overlap distribution becomes independent of the system syze in thermodynamic
limit $N\rightarrow \infty $ . The asymptotic overlap distribution is formed
by a delta-peak at $q=1$ plus a broad, smooth part extending to low values
of the overlap $q$. The presence of such continuous distribution is an
indication of replica-symmetry breaking. The breakdown of replica symmetry
means that, for \textit{each} orbit of GCLM, one can find orbits of
gradually varying degrees of similarity within a large ensemble of orbits
generated by randomly chosen initial conditions.

\subsection{Ultrametricity}

Generally, the ultrametric distance $d(A,B)$ between two elements $A$ and $B$
in a hierarchy is defined as the number of steps one should go up in the
hierarchy to find a common ancestor of two elements $A$ and $B$. If any
three elements $A,$ $B$ and $C$ belong to a hierarchy, the inequality $%
d(A,C)\leq \max \{d(A,B),d(B,C)\}$ should hold. As a consequence, the two
maximal distances between elements in any triad must always be equal. If
overlaps $q^{\alpha \beta }$ between any two replicas $\alpha $ and $\beta $
are uniquely determined by the ultrametric distance $d(\alpha ,\beta )$
between the respective states, the overlaps between any three replicas $%
\alpha ,\beta $ and $\gamma $ must satisfy the relationship 

\begin{equation} 
\label{ultraeq}
q^{\alpha \gamma }\geq \min \{q^{\alpha \beta },q^{\alpha \gamma }\} \, ,
\end{equation}
implying that
the two minimal overlaps in any triad of replicas are always equal 
\cite{RMP}.

To check the presence of ultrametricity, we have to consider triads of
replicas $\alpha $, $\beta $, and $\gamma $ and calculate the three overlaps
that can be defined by combining them. If the two minimal overlaps in any
triad are always equal, the ultrametricity is present. Formally, this
amounts to requiring that the relationship (\ref{ultraeq}) always holds. That 
condition can
be numerically tested by generating triads of replicas and computing the
distribution $H(\Delta q)$ over the differences between the two minimal
overlaps, $\Delta q\equiv |q^{\alpha \beta }-q^{\alpha \gamma }|$. If $%
H(\Delta q)\to \delta (\Delta q)$ in the limit $N\to \infty $, then the
system is ultrametric.

\begin{figure}[tbp]
\resizebox{0.45\textwidth}{!}{\includegraphics{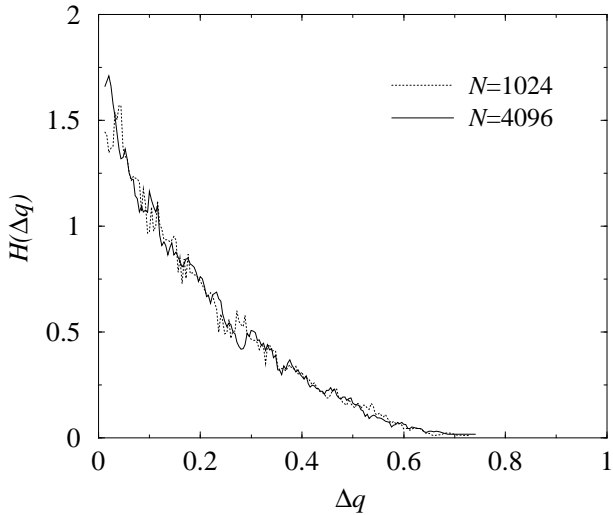}}
\caption{Distribution $H(\Delta q)$ of distances between the two smaller
overlaps our of a triad. In the thermodynamic limit this function has a
non-vanishing width. Same parameters as in Fig.~15.}
\label{Fig:contult}
\end{figure}

Previously, such calculations have been performed for the fixed-field
ensemble \cite{MM2}. In this case case the distribution $H(\Delta q)$
approaches a delta-function $\delta (\Delta q)$ in the limit of large system
sizes $N$. However, as becomes clear from the analysis of overlap
distributions in the present study, this behaviour simply reflects the
vanishing diversity of system attractors for the fixed-field ensemble in the
thermodynamic limit.

We have now repeated such calculations for the random-field ensemble.
Distributions $H(\Delta q)$ of distances between the two smaller overlaps in
randomly generated triads of replicas for two different system sizes are
shown in Fig. 16. We see that the distributions are broad and almost do not
depend on the system size. Thus, GCLM do not display ultrametric properties.

Though replica-symmetry breaking is a necessary condition for nontrivial
overlap distributions, it does not imply exact ultrametricity, which is a much
more demanding condition. Possible
deviations from exact ultrametricity have been discussed for spin glasses
\cite{RMP}. Parisi and Ricci-Tersenghi \cite{Parisi99} have shown that
exact ultrametricity can only hold under the conditions of {\it stochastic
stability} (i.e. that each replica is in a certain sense equivalent to the
others) and of {\it separability} (i.e. that all the mutual information about
a pair of equilibrium configurations is already encoded in their overlap).
Our numerical analysis of GCLM shows that for the random-field ensemble in
the thermodynamical limit this system is characterized by replica-symmetry
breaking, but exact ultrametricity is absent. Note that though exact
ultrametricity, which would have corresponded to the appearance of
delta-function peak at $\Delta q=0$, is not observed, the distributions $%
H(\Delta q)$ in Fig.~16 have a broad maximum at $\Delta q=0$. This indicates
that some weaker form of organization may still be present here.

\section{Discussion and conclusions}

We have examined asymptotic glass properties of globally coupled logistic
maps in the thermodynamic limit for two different random ensembles of
initial conditions. In the fixed-field ensemble the initial value of the
synchronizing field becomes identical up to variations of order $1/\sqrt{N}$
that vanish in the thermodynamic limit.Therefore, all attractors reached by
the system become increasingly similar for $N\rightarrow \infty $.
Dynamically, the bias due to the initial field drives the system towards the
prefered attractor and then the elements partition in such a way to
stabilize the prefered attractor. The overlap distribution tends to a
delta-function peak at $q=1$, i.e., even when a diverging number of
attractors is present, they are all macroscopically identical. Hence,
replica symmetry is recoverd for the fixed-field ensemble in the
thermodynamic limit.

We have also found that, in the fixed-field ensemble, the system undergoes a
special phase transition when $\epsilon $ overcomes a critical value. For
smaller coupling, the system is partitioned into a small number (close to
the transition, usually two) of periodic orbits. Though all of the
attractors reached for different initial conditions are very similar, their
number diverges in the thermodynamic limit, and their average attraction
basin weight goes to zero. For couplings larger than the critical one, the
system synchronizes completely for nearly all initial conditions in the
thermodynamic limit, and the average attraction basin weight tends to unity.
This situation is reminiscent to the analogous transition in attraction
basin weights observed for random boolean networks \cite{Bast1} and for
asymmetric neural networks \cite{Bast2}. In both cases, the average
attraction basin goes discountinuously (in the thermodynamic limit) from the
value zero, when the system is in the ``ordered phase'', to a finite value
related to the average attraction basin of random maps \cite{REM} in the
chaotic phase. Though in GCLM the finite attraction basin weight is a
consequence of complete synchronization, the formal analogy between this
system and dynamical systems with quenched disorder is very suggestive.

The asymptotic behaviour of GCLM in the thermodynamic limit is essentially
different when the random-field ensemble of initial conditions is chosen.
Because initial synchronizing fields retain in this case macroscopic
fluctuations even for $N\rightarrow \infty $, a broad range of attractors
may still be reached. In the random-field ensemble, the transition to
complete synchronization (with the average attraction basin weight $Y=1$) is
expected to be continuous, more in analogy with equilibrium mean-field spin
glasses. Examination of the overlap distributions has revealed that
replica-symmetry breaking persists in the thermodynamic limit. Thus, GCLM reach
the status of a dynamical countepart to mean-field spin glasses.


\begin{thebibliography}{99}
\bibitem{Kan}  K. Kaneko, Physica D \textbf{41}, (1990) 137-172.

\bibitem{Kan1}  K. Kaneko, Phys. Rev. Lett. \textbf{63}, (1989) 219-223.

\bibitem{Kbook}  K. Kaneko and I. Tsuda, \textit{Complex systems: Chaos and
Beyond} (Springer-Verlag, Berlin Heidelberg 2001).

\bibitem{MM1}  S.C. Manrubia and A.S. Mikhailov, Europhys. Lett. \textbf{50}
, (2000) 580-586.

\bibitem{Abramson}  G. Abramson, Europhys. Lett. \textbf{52}, (2000) 615-619.

\bibitem{Shibata}  T. Shibata, T. Chawanya, and K. Kaneko, Phys. Rev. Lett.
\textbf{82}, (1999) 4424-4427.

\bibitem{Chaos}  N.J. Balmforth, A. Jacobson, and A. Provenzale, Chaos
\textbf{9}, (1999) 738-754.

\bibitem{Mezard}  M. M\'{e}zard, G. Parisi, and M.A. Virasoro, \textit{\
Spin-Glasses Theory and Beyond} (World Scientific, Singapore 1987).

\bibitem{Kanglass}  K. Kaneko, J. Phys. A \textbf{24} (1991) 2107-2119;
Physica D \textbf{124}, (1998) 322-344.

\bibitem{Vulpiani}  A. Crisanti, M. Falcioni, and A. Vulpiani, Phys. Rev.
Lett. \textbf{76}, (1996) 612-615.

\bibitem{Ksplit}  K. Kaneko, Physica D \textbf{77}, (1994) 456-472.

\bibitem{MM2}  S.C. Manrubia and A.S. Mikhailov, Europhys. Lett. 
\textbf{53}, (2001) 451-457.

\bibitem{RMP} R. Rammal, G. Toulouse, and M.A. Virasoro, Rev. Mod. Phys. 
{\bf 58}, (1986) 765-.

\bibitem{Parisi99}  G. Parisi and F. Ricci-Tersenghi, J. Phys. A {\bf 33},
(2000) 113-.

\bibitem{Fischer}  K.H. Fischer and J.A. Hertz, \textit{Spin Glasses}
(Cambridge University Press, Cambridge 1991).

\bibitem{Bast1}  U. Bastolla and G. Parisi, Physica D \textbf{115}, (1998)
203-218; \textit{ibid.} \textbf{115}, (1998) 219-233.

\bibitem{Bast2}  U. Bastolla and G. Parisi, J. Phys. A: Math. Gen. \textbf{30%
}, (1997) 5613-5631; \textit{ibid.} \textbf{31}, (1998) 4583-4602.

\bibitem{REM}  B. Derrida and H. Flyvbjerg, J. de Physique \textbf{48},
(1986) 971-978.

\end{thebibliography}
\end{document}